\documentclass[10pt]{article}
\usepackage{graphicx}

\def\Title#1{\begin{center} {\Large #1 } \end{center}}
\def\Author#1{\begin{center}{ \sc #1} \end{center}}
\def\Address#1{\begin{center}{ \it #1} \end{center}}

\newcommand\pubblock{\rightline{\begin{tabular}{l} Proceedings of the Fifth Annual LHCP\\ \pubnumber\\
         \pubdate  \end{tabular}}}

\newenvironment{Abstract}{\begin{quotation} \begin{center} 
             \large ABSTRACT \end{center}\bigskip 
      \begin{center}\begin{large}}{\end{large}\end{center} \end{quotation}}

\newenvironment{Presented}{\begin{quotation} \begin{center} 
             PRESENTED AT\end{center}\bigskip 
      \begin{center}\begin{large}}{\end{large}\end{center} \end{quotation}}

\def\beq{\begin{equation}}
\def\eeq#1{\label{#1}\end{equation}}
\def\eeqn{\end{equation}}

\def\beqa{\begin{eqnarray}}
\def\eeqa#1{\label{#1}\end{eqnarray}}
\def\eeqan{\end{eqnarray}}

\let\bar=\overbar

\def\Dslash{\not{\hbox{\kern-4pt $D$}}}
\def\dslash{\not{\hbox{\kern-2pt $\del$}}}

\def\msb{{\bar{\ssstyle M \kern -1pt S}}}

\usepackage{color}
\usepackage{caption}
\usepackage{topcapt}
\usepackage[caption=false]{subfig}
\let\subfigure\subfloat
 \usepackage{float}
 
\newcommand\pubnumber{ CMS CR-2017/215 }
\newcommand\pubdate{\today}

\def\affiliation{
On behalf of the CMS Collaboration, \\
Laboratoire Leprince-Ringuet \\
\'{E}cole polytechnique, 91120 Palaiseau, France}

\usepackage{xspace}
\newcommand{\GeV}{\ensuremath{\,\mathrm{Ge\hspace{-.08em}V}}\xspace}
\newcommand{\TeV}{\ensuremath{\,\mathrm{Te\hspace{-.08em}V}}\xspace}

\newcommand{\pt}{\ensuremath{p_{\mathrm{T}}}\xspace}
\newcommand{\kt}{\ensuremath{k_{\mathrm{T}}}\xspace}

\newcommand{\fb} {\mbox{\ensuremath{\,\mathrm{fb}}}\xspace}

\newcommand{\stat}{\ensuremath{\,\mathrm{(stat)}}\xspace}
\newcommand{\syst}{\ensuremath{\,\mathrm{(syst)}}\xspace}

\begin{document}

% large size for the first page
\large
\begin{titlepage}
\pubblock

%% Change the title, name, abstract
%% Title 
\vfill
\Title{Vector boson fusion and scattering results from CMS}
\vfill

%  if you need to add the support use this, fill the \support definition above. 
%   \Author{ FIRSTNAME LASTNAME \support }
\Author{ Philipp Pigard  }
\Address{\affiliation}
\vfill
\begin{Abstract}

Measurements on vector boson scattering in the same-sign WW and fully leptonic ZZ channel, and a measurement of vector boson fusion of a Z boson are summarized. The three measurements are based on a dataset of proton--proton collisions at $\sqrt{s}=13\,\mathrm{Te\hspace{-.08em}V}$ with an integrated luminosity of $35.9\mbox{\ensuremath{\,\mathrm{fb}}}^{-1}$ recorded by the CMS experiment. The first observation for the electroweak production of a pair of same-sign W bosons is reported with an observed (expected) significance of 5.5 (5.7) standard deviations. The first measurement of vector boson fusion of a Z boson at $\sqrt{s}=13\,\mathrm{Te\hspace{-.08em}V}$ is presented. The first measurement of the electroweak production of two Z bosons at the LHC is also presented, reporting a signal significance of 2.7 standard deviations (1.6 standard deviations expected). The data are in general agreement with the expectations from the standard model. Limits on physics beyond the standard model are presented in the vector boson scattering analyses, providing the most stringent constraints on such scenarios to date.

\end{Abstract}
\vfill

% DO NOT CHANGE 
\begin{Presented}
The Fifth Annual Conference\\
 on Large Hadron Collider Physics \\
Shanghai Jiao Tong University, Shanghai, China\\ 
May 15--20, 2017
\end{Presented}
\vfill
\end{titlepage}
\def\thefootnote{\fnsymbol{footnote}}
\setcounter{footnote}{0}
%

% normal size for the rest
\normalsize 

%% Your paper should be entered below. 

\section{Vector boson scattering and experimental status}
Vector boson scattering (VBS) and vector boson fusion (VBF) processes are central predictions of the standard model (SM) and a direct consequence of the non-Abelian gauge structure of the electroweak (EW) interaction. The VBS process class furthermore permits to investigate the breaking of the EW symmetry and enables complementary studies of the Higgs boson discovered by the ATLAS and CMS Collaborations~\cite{Aad:2012tfa,Chatrchyan:2012xdj}. Without further contributions, the scattering amplitudes for the longitudinal polarizations of the weak gauge bosons would violate unitarity at scattering energies of about 1\TeV~\cite{Lee:1977yc,Lee:1977eg}. In the minimal scalar sector of the SM, the divergent behaviour of the longitudinal scattering amplitudes is cancelled by the  interference between these pure-gauge amplitudes and amplitudes that feature the Higgs boson, provided the Higgs-gauge (HVV) couplings are as prescribed by the Brout-Englert-Higgs mechanism. The study of differential cross sections of VBS at large scattering energies permits to constrain the HVV couplings, complementing the measurements of Higgs boson production and decay rates. The topology of VBS processes furthermore facilitates the search for beyond the SM effects, both for particular models or in a generic effective field theory approach of anomalous quartic gauge couplings~(aQGCs)~\cite{Degrande:2012wf, Eboli:2006wa}.
%~\cite{Khachatryan:2016vau}

At the LHC, VBF and VBS are initiated by quarks q from the colliding protons. Each quark radiates a weak boson, which then fuse (VBF, $2\rightarrow1$) or scatter (VBS, $2\rightarrow2$). The outgoing quarks give rise to hadronic jets commonly referred to as tagging jets. These tagging jets in opposite hemispheres of the detector and at large pseudorapidities, resulting in large tagging jet pseudorapidity separations and invariant masses, are the hallmark signatures of VBF and VBS processes. The hard interaction in both processes involves only the EW interaction, which suppresses color exchange between the quarks and results in a general suppression of additional hadronic activity aside from the tagging jets. 
The theoretical modeling of such extra emissions can be studied in either VBF or VBS processes. However, VBF processes are of particular importance as they feature considerably larger cross sections.

A common background to VBF and VBS processes arises when the observed final state is produced via quantum chromodynamics (QCD). In the absence of specific circumstances, these processes constitute an important source of noninstrumental background in VBF and VBS studies.

This proceeding reports on three recent measurements of VBF and VBS by the CMS Collaboration~\cite{CMS-PAS-SMP-17-004, CMS-PAS-SMP-16-018, ZZjj}. The measurements exploit a dataset of proton--proton collisions collected in 2016 at a center-of-mass energy of $\sqrt{s}=13\TeV$ with an integrated luminosity of $35.9\fb^{-1}$. A detailed description of the CMS apparatus and the coordinate system is given in Ref.~\cite{CMS_detector}. The collision events are reconstructed by combining the measurements of the subdetectors in a particle-flow (PF) algorithm which returns a list of PF candidates. Jets are reconstructed using PF candidates and the anti-\kt clustering algorithm with a distance parameter $R = 0.4$.

\section{VBS in the same-sign WW channel}

The same-sign WW channel of VBS, where the final state is characterized by two prompt leptons of opposite charge, moderate missing transverse momentum and the two tagging jets, provides a favorable experimental signature. The same-charge requirement severely reduces the contributions of the QCD background, the leptons provide a clear signature for triggering and the event selection, and it features the largest cross section of any VBS process at the LHC. The CMS Collaboration performed a first measurement of this process at $8\TeV$ \cite{CMS_ssWW}. The results presented here are reported in Ref.~\cite{CMS-PAS-SMP-17-004}.

The event selection requires two same-sign lepton candidates, muons or electrons, with $\pt > 25\,(20)\GeV$ for the \pt-leading (subleading) lepton and $|\eta| < 2.4\,(2.5)$ for muons (electrons). Electrons and muons are required to be isolated from other charged and neutral particles in the event and are required to satisfy quality criteria.
Events are required to have at least two selected jets with $\pt > 30\GeV$ and $|\eta| < 5.0$. The two \pt-leading jets are referred to as the tagging jets. A VBS selection is defined by requiring  a large dijet mass ($m_\mathrm{jj} > 500\GeV$), large pseudorapidity separation ($|\Delta\eta_\mathrm{jj}| > 2.5$), and $\mathrm{max}(z^*) < 0.75$, where $z^* = |\eta_\ell - (\eta_\mathrm{j1} + \eta_\mathrm{j2})/2|/ |\Delta\eta_\mathrm{jj}|$. 

Additional selections help to reduce backgrounds. Backgrounds from processes featuring a top-quark decay are suppressed by vetoing events via a bottom-quark tagging technique. The dilepton invariant mass is required to satisfy $m_{\ell\ell} > 20\GeV$, the missing transverse momentum is required to exceed 40\GeV. To suppress the background from charge misidentication in Drell-Yan events, dielectron events need to satisfy $|m_{\ell\ell} - m_\mathrm{Z}| > 15\GeV$. Finally, WZ events featuring a third, loosely identified lepton (hadronic tau decay) with $\pt > 10\,(18)\GeV$ are vetoed.

\begin{figure}[H] \centering
\captionsetup[subfigure]{labelformat=empty}
\subfigure [] {\resizebox{0.4 \textwidth}{!}{\includegraphics{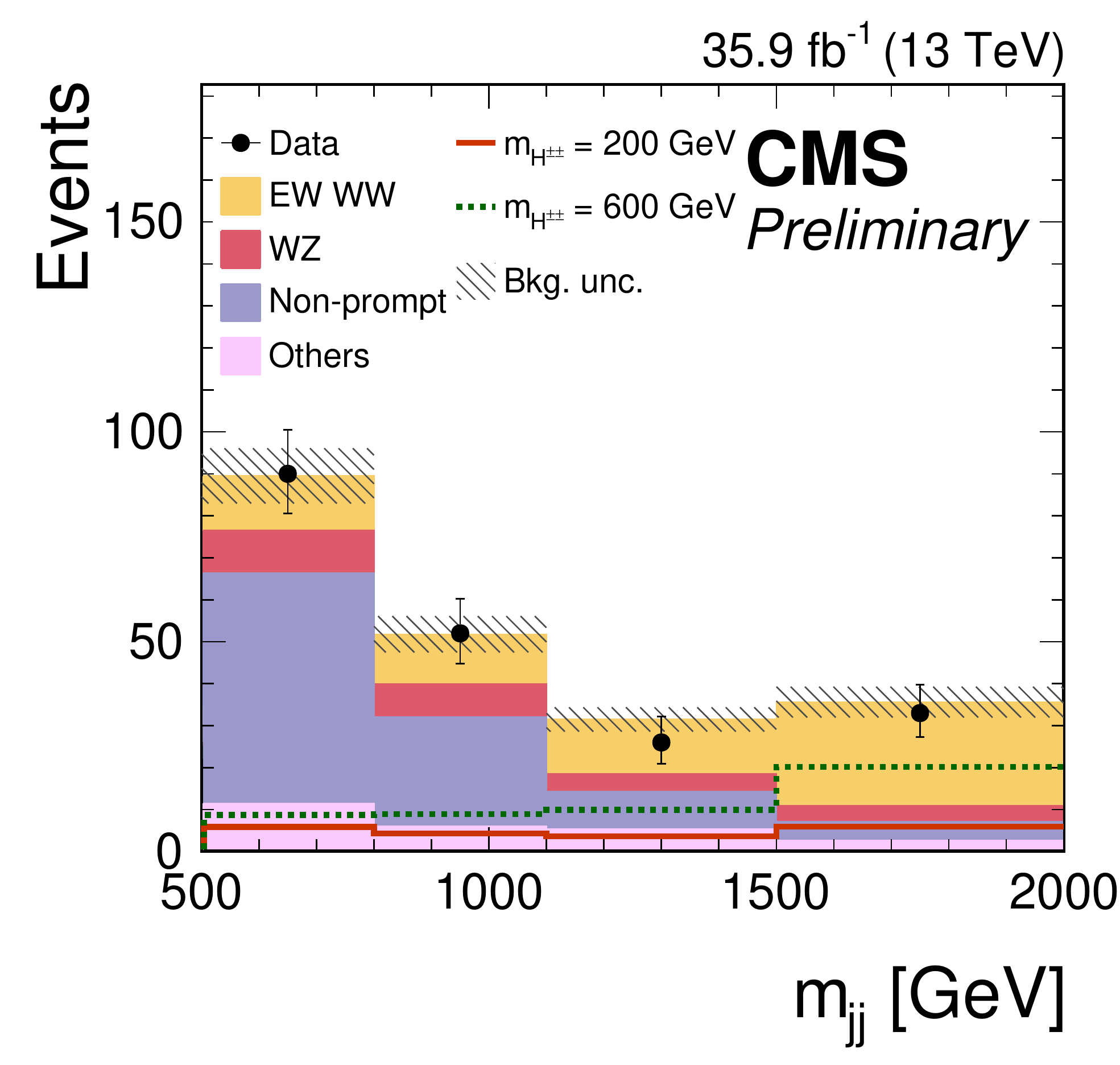}}}
\subfigure [] {\resizebox{0.4 \textwidth}{!}{\includegraphics{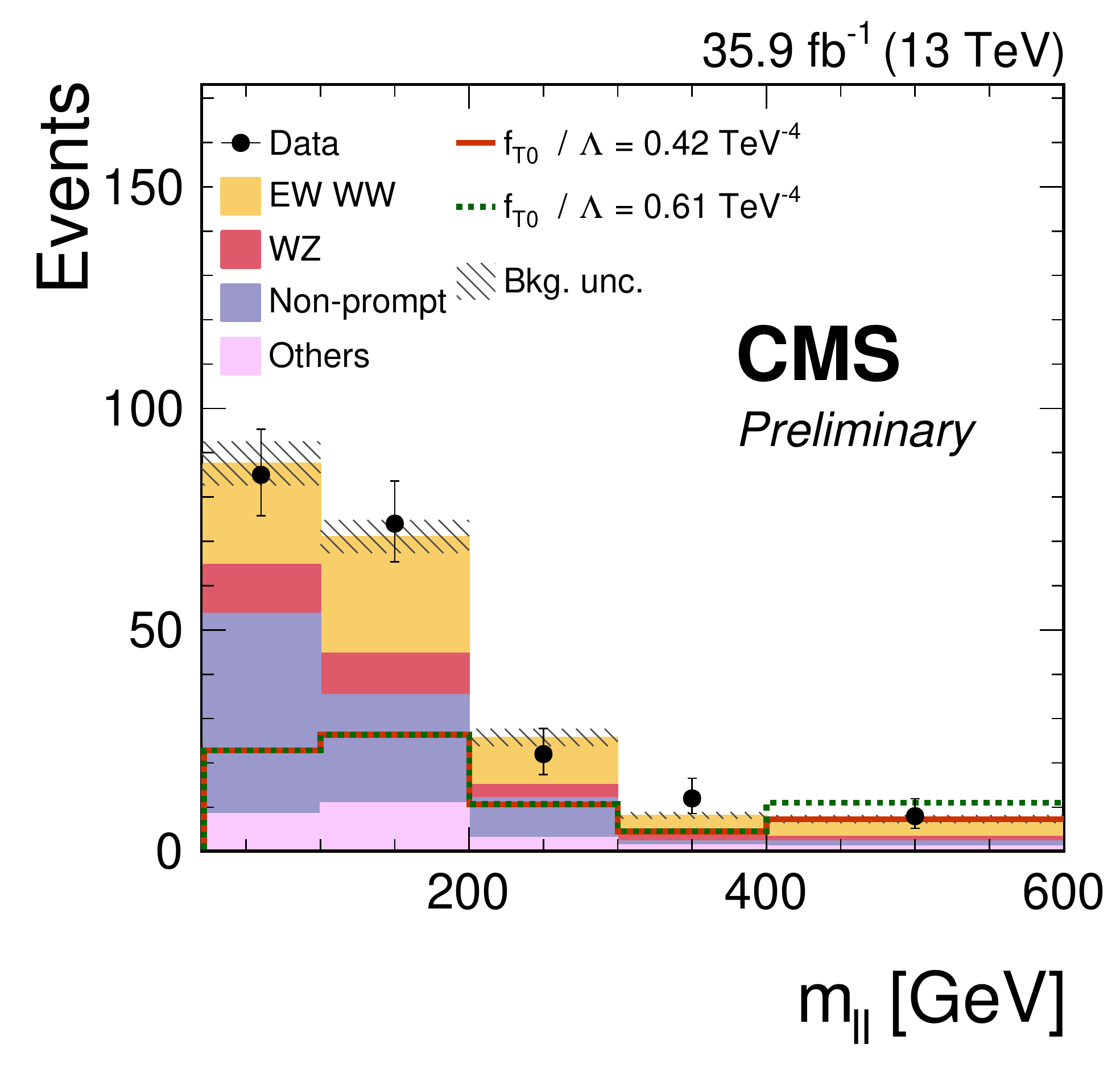}}}
\caption{Distribution of the tagging jet invariant mass (left panel) and dilepton invariant mass (right panel)~\cite{CMS-PAS-SMP-17-004}. The last bins include the contributions from $m_\mathrm{jj}>1500\GeV$ and $m_{\ell\ell}>400\GeV$ respectively and the hatched bands include the statistical and systematic uncertainties. The left panel also shows the expected yields from two hypotheses of charged Higgs boson production, while the right panel shows two aQGC hypotheses.
}\label{ssWW_1}
\end{figure}

The two dominating sources of background events after this event selection arise from nonprompt leptons and WZ. Both backgrounds are estimated from data using control regions. Other background processes are estimated using simulation. Figure~\ref{ssWW_1} shows the observed $m_{\ell\ell}$ and $m_\mathrm{jj}$ distributions in the data and the expectations from the signal plus background model. An excess above the background-only hypothesis is observed in the data, with an observed (expected) significance of 5.5 (5.7) standard deviations . The signal strength $\mu$, defined as the ratio between the expected number of signal events from the SM prediction and the observed number of events, is $\mu=0.90\pm0.22$. A fiducial cross section is reported as 
$\sigma_\mathrm{fid}(\mathrm{W}^\pm\mathrm{W}^\pm\mathrm{jj}) = 3.83\pm0.66\stat\,\pm0.35\syst\fb$, in agreement with the SM expectation of $4.25 \pm 0.21$.

Certain models of beyond the SM physics predict doubly-charged Higgs boson states, and the selected same-sign events are used to constrain the parameters of such models. Finally, the data is used to constrain aQGC couplings, greatly improving previous confidence limits (CL) as summarized in Tab.~\ref{tab:VBS_aQGC}.

\begin{table}[h!]
    \centering
      \begin{tabular}{lccc}
      \hline \hline
                          & Observed limits  & Expected limits  & Run-I limits \\
                          & (\TeV$^{-4}$)   & (\TeV$^{-4}$)   & (\TeV$^{-4}$) \\
        \hline 
          $\mathrm{f_{S0}} / \Lambda$  & [ -7.7, 7.7] & [ -7.0, 7.2] & [-38 , 40]\\ 
          $\mathrm{f_{S1}} / \Lambda$  & [-21.6,21.8] & [-19.9,20.2] & [-118 , 120]\\ 
          $\mathrm{f_{M0}} / \Lambda$  & [ -6.0, 5.9] & [ -5.6, 5.5] & [-4.6 , 4.6] \\ 
          $\mathrm{f_{M1}} / \Lambda$  & [ -8.7 ,9.1] & [ -7.9, 8.5] & [-17 , 17] \\ 
          $\mathrm{f_{M6}} / \Lambda$  & [-11.9,11.8] & [-11.1,11.0] & [-65 , 63]   \\ 
          $\mathrm{f_{M7}} / \Lambda$  & [-13.3,12.9] & [-12.4,11.8] & [-70 , 66] \\ 
          $\mathrm{f_{T0}} / \Lambda$  & [-0.62,0.65] & [-0.58,0.61] & [-3.8 , 3.4]\\ 
          $\mathrm{f_{T1}} / \Lambda$  & [-0.28,0.31] & [-0.26,0.29] & [-1.9 , 2.2]\\ 
          $\mathrm{f_{T2}} / \Lambda$  & [-0.89,1.02] & [-0.80,0.95] & [-5.2 , 6.4] \\ 
       \hline \hline
      \end{tabular}
    \caption{
         Observed and expected 95\% CL limits on aQGC coupling parameters as reported in \cite{CMS-PAS-SMP-17-004}. The last column summarizes the LHC Run-I observed limits obtained by CMS. 
       }
    \label{tab:VBS_aQGC}
\end{table}

\section{VBF with a Z boson}
The electroweak production of a Z boson in conjunction with two jets (Zjj) is an important test of the SM and permits to  study central jet activity of VBF-type processes. The CMS Collaboration performed measurements of VBF in Zjj at 7 and 8\TeV~\cite{Chatrchyan:2013jya, Khachatryan:2014dea}. The measurement at 13\TeV is reported in Ref.~\cite{CMS-PAS-SMP-16-018}.

The analysis selects leptonic Z boson decays, i.e., events that feature a pair of opposite-charge, same-flavor leptons. Reconstructed electrons (muons) are required to satisfy $|\eta^\ell|<2.5\,(2.4)$ and the lepton with highest (next-to-highest) \pt is required to be $\pt>30\,(20)\GeV$. Events with same-flavor dileptons (ee or $\mu\mu$) and $|m_\mathrm{Z} - m_{\ell\ell}| < 15\GeV$ are selected. The \pt-leading (subleading) jets inside the range $|\eta|<4.7$ are required to satisfy $\pt>50\,(30)\GeV$ as well as $m_\mathrm{jj}>200\GeV$ and are referred to as the tagging jets.

The dominant background in this analysis is the Drell-Yan plus jets process. The VBF signal is extracted via a boosted decision tree (BDT) discriminant that exploits the tagging jet invariant mass $m_\mathrm{jj}$, pseudorapidity separation $|\Delta\eta_\mathrm{jj}|$, the centrality of the Z boson $z^* = |\eta_\mathrm{Z}- (\eta_\mathrm{j1} + \eta_\mathrm{j2})/2|/ |\Delta\eta_\mathrm{jj}|$, and the output of a quark-gluon discriminator. Two observables sensitive to the angular distributions are also considered in the BDT: the ratio  between the \pt of the tagging jet system and the scalar \pt sum of the tagging jets ($R\pt^\mathrm{jet}$), and $R\pt^\mathrm{hard}$, which is defined as the transverse component of the vector sum of the Z boson and the tagging jet momenta, normalized to the scalar \pt sum of these objects.

\begin{figure}[H]  \centering
\subfigure [] {\resizebox{0.4 \textwidth}{!}{\includegraphics{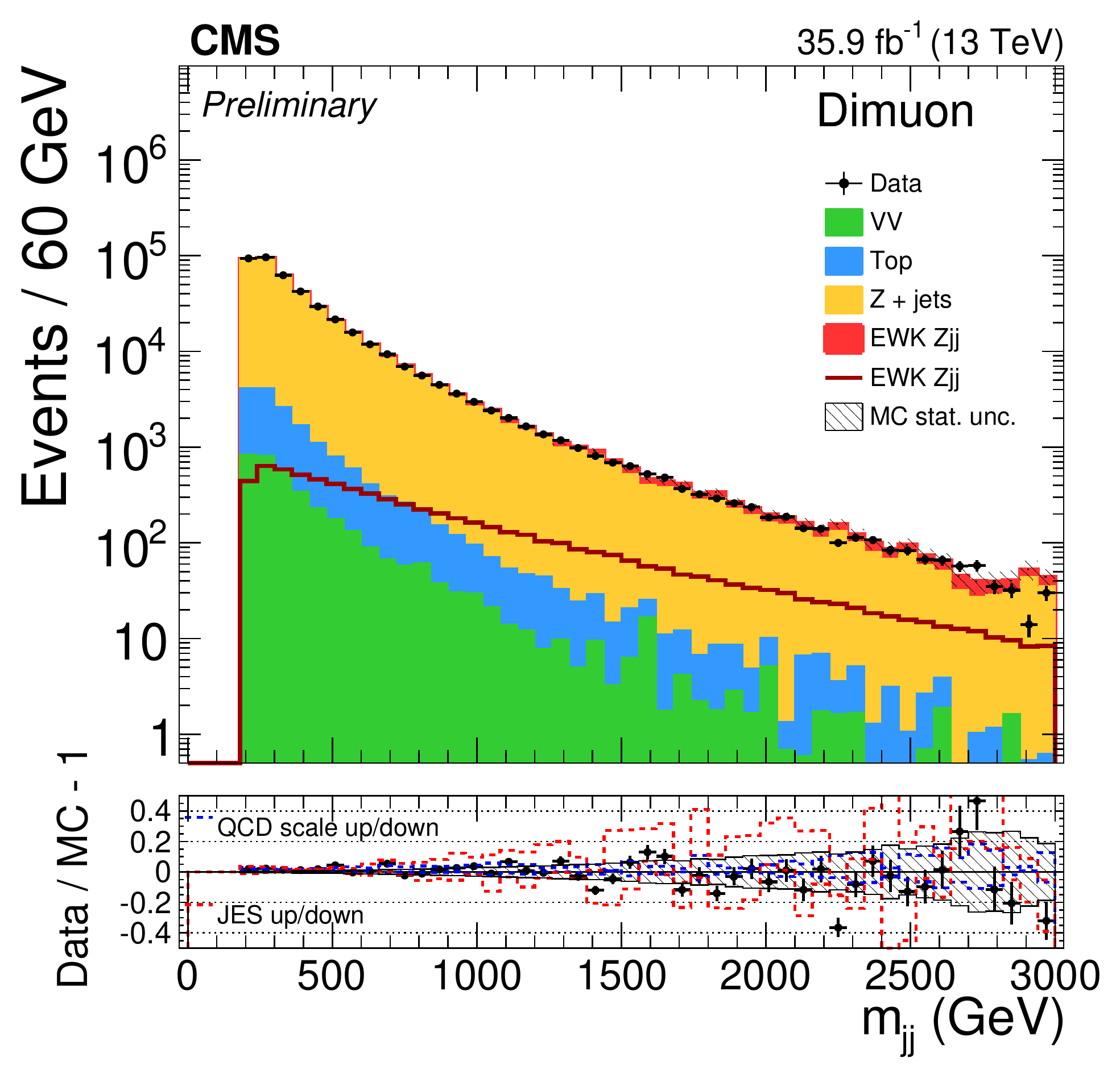}}}
\subfigure [] {\resizebox{0.4\textwidth}{!}{\includegraphics{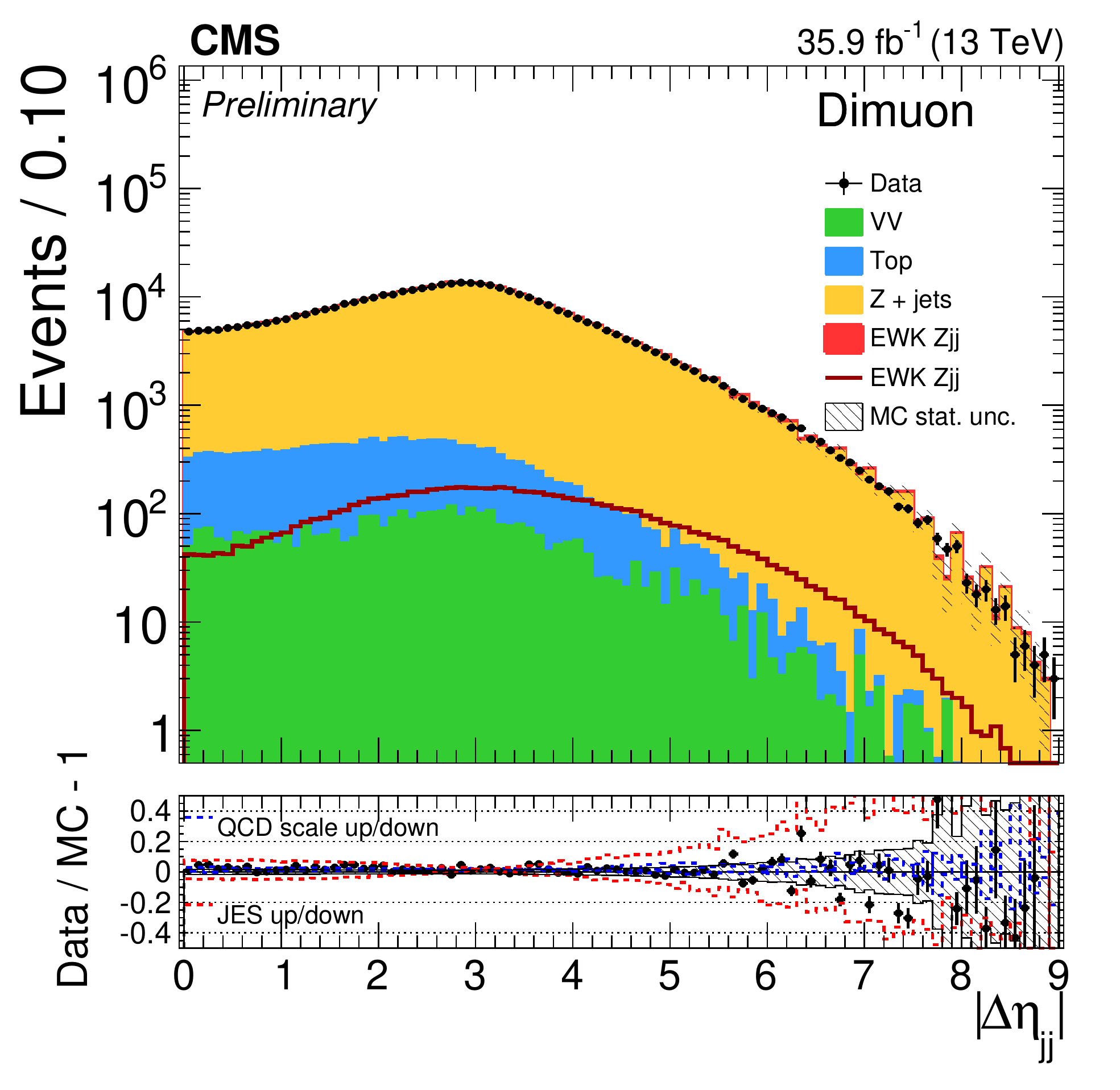}}}

\caption{Distribution of the tagging jet invariant mass (left panel) and tagging jet pseudorapidity separation (right panel) for Zjj events observed in the data~\cite{CMS-PAS-SMP-16-018}.
}\label{VBF_1}
\end{figure}

\begin{figure}[H] \centering
\subfigure [] {\resizebox{0.4 \textwidth}{!}{\includegraphics{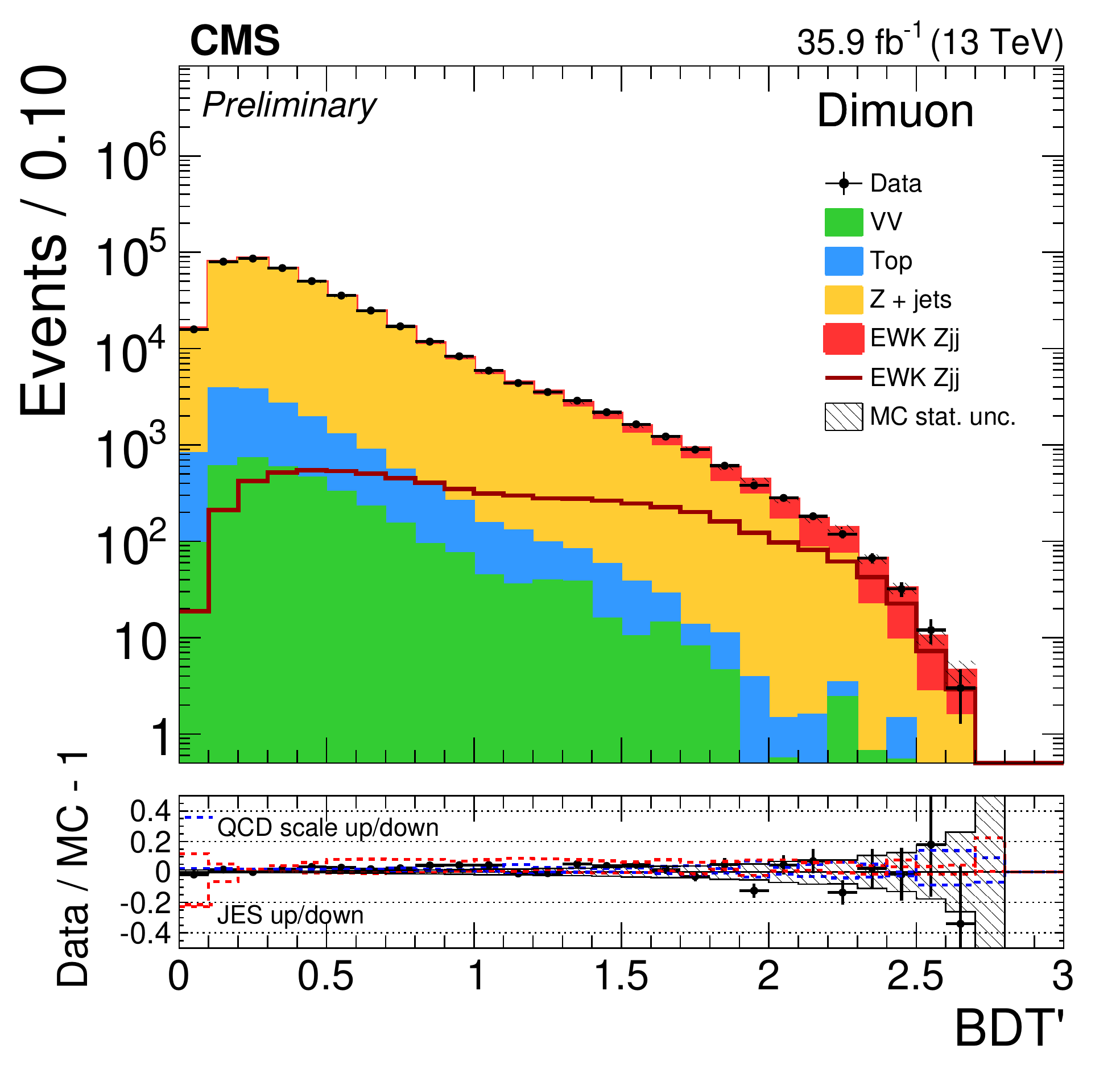}}}
\subfigure [] {\resizebox{0.4 \textwidth}{!}{\includegraphics{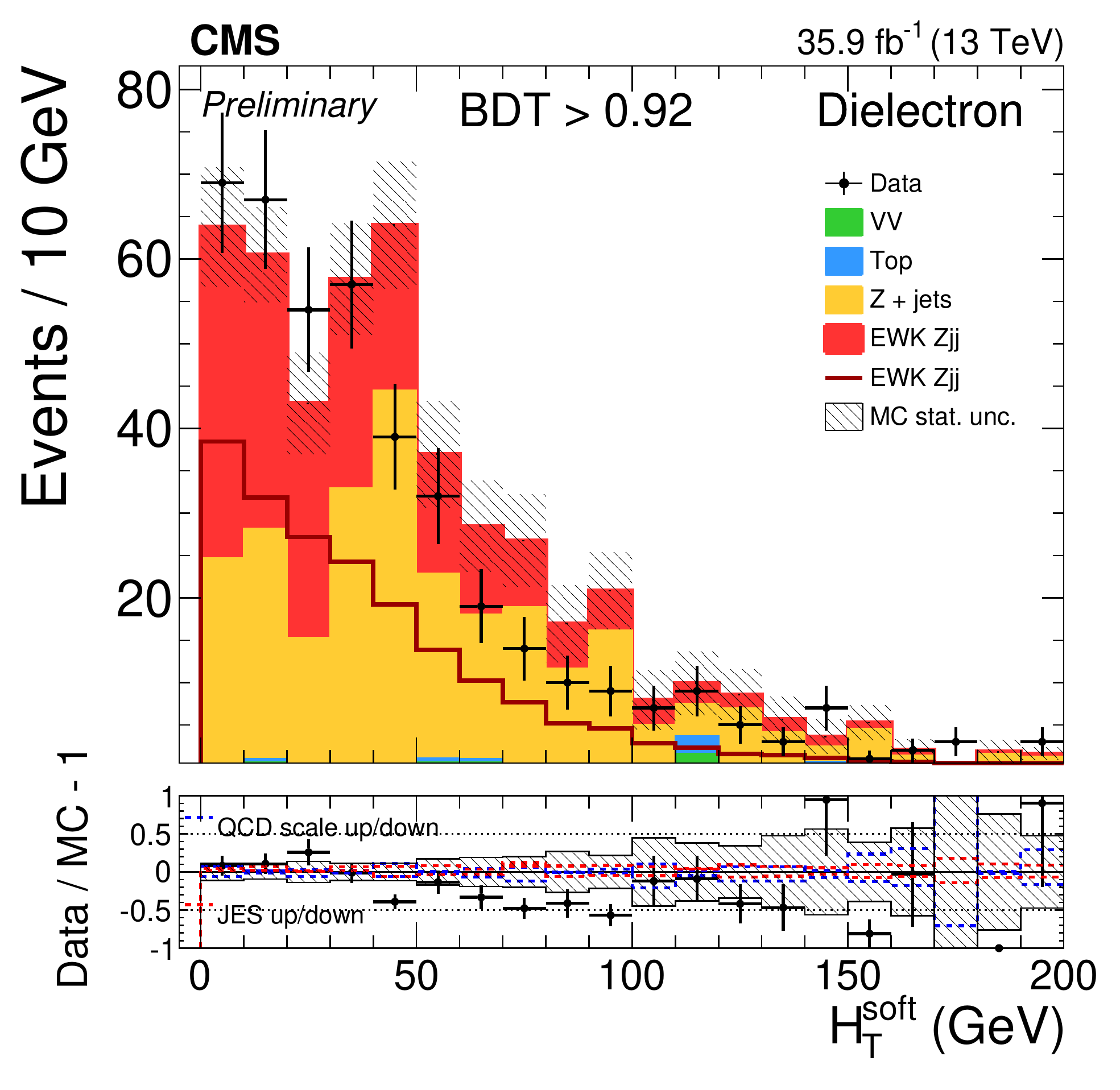}}}
\caption{Distribution of the BDT output observed the data (left panel) and hadronic activity for VBF-like events (right panel) \cite{CMS-PAS-SMP-16-018}.
}\label{VBF_2}
\end{figure}

The left (right) panel of Fig.~\ref{VBF_1} shows the distributions of the tagging jet invariant mass $m_\mathrm{jj}$ (pseudorapidity separation $|\Delta\eta_\mathrm{jj}|$). The BDT separates the Drell-Yan background from the EW Zjj signal, which predominates at large BDT output values, as shown in the left  panel of Fig.~\ref{VBF_2}. The BDT output distribution is used to measure the signal strength for the EW production of Zjj by fitting the signal plus background model obtained from the simulation to the data. The signal is established at more than 5 standard deviations and the observed signal strength of a simultaneous  measurement in the dielectron and dimuon channels is 
$\mu = 1.017 \pm 0.035\stat\pm 0.101\syst = 1.017\pm0.106\,(\mathrm{total})$. 

A fiducial cross section is measured as $\sigma(\mathrm{EW}~\ell^+\ell^-\mathrm{jj}) = 552\pm19\stat\pm 55\syst\fb = 552\pm58\,(\mathrm{total})\fb$, compatible with the SM prediction of $543\pm24\fb$.

The BDT is furthermore used to select VBF-like events and to study the presence and kinematics of additional hadronic activity. The scalar \pt sum of all charged tracks with $\pt>1\GeV$ is used to define the $\mathrm{H}_\mathrm{T}^\mathrm{soft}$ observable. The right panel of Fig.~\ref{VBF_2} compares the distribution of this observable in the data with a parton-shower prediction.

\section{VBS in the ZZ channel}

A measurement of VBS in the ZZ channel is presented, where both Z bosons are identified by their decays to electrons or muons, making for a clean experimental signature. The fully leptonic final state of the ZZ channel is the only VBS process that enables the exact reconstruction of all final state particles, which gives access to the vector boson scattering energy and the decay angles of the leptons permits to select extraction of the longitudinal contribution to VBS. However, a small cross section, the small $\mathrm{Z}\rightarrow\ell\ell$ branching fraction, and a large background from QCD-induced production of the same final state result in a low signal-over-background ratio. Ref.~\cite{ZZjj} presents the first measurement of VBS in the ZZ channel at the LHC\footnote{The measurement was first made public as part of CMS Physics Analysis Summary CMS-PAS-16-019. The figures presented in this proceeding are those of Ref.~\cite{ZZjj}, with minor visual modifications.}.

The event selection aims to identify events with four leptons from the decay of two Z bosons and two jets. Events are recorded for analysis by single, dilepton, and trilepton triggers for an overall trigger efficiency larger than 98\%. Z boson candidates are constructed from opposite-sign, same-flavor dilepton pairs, from which ZZ candidates are built. The \pt-leading  and \pt-subleading muon (electron) in a ZZ candidate has to satisfy $\pt>17\,(23)$ and $\pt>8\,(12)$, respectively. The final ZZ candidate is determined based on the maximum scalar sum of the lepton \pt. Events are selected if the Z boson candidates satisfy $60 < m_{\ell\ell}<120\GeV$. The tagging jets are chosen as the two \pt-leading jets with $\pt^\mathrm{j}>30\GeV$ and $|\eta^\mathrm{j}|<4.7$. Events that feature a ZZ candidate and tagging jets that satisfy $m_\mathrm{jj}>100\GeV$ are considered in the analysis.

The contribution of the EW production to all selected ZZjj events is 5\%, with more than 80\% of the expected yield originating from the noninstrumental background via QCD-induced production of the ZZjj signature. A multivariate discriminant based on seven observables is used to separate the EW signal from the background. The BDT exploits the tagging jet invariant mass $m_\mathrm{jj}$, the pseudorapidity separation between the tagging jets $|\Delta\eta_\mathrm{jj}|$, the invarant mass of the ZZ system $m_\mathrm{ZZ}$, the $z^*$ observables of both Z bosons, and  the event balance observables $R\pt^\mathrm{hard}$ and $R\pt^\mathrm{jets}$. The separation power of this BDT is found to be identical to a matrix element based approach. 

\begin{figure}[H] \centering
\subfigure [] {\resizebox{0.4 \textwidth}{!}{\includegraphics{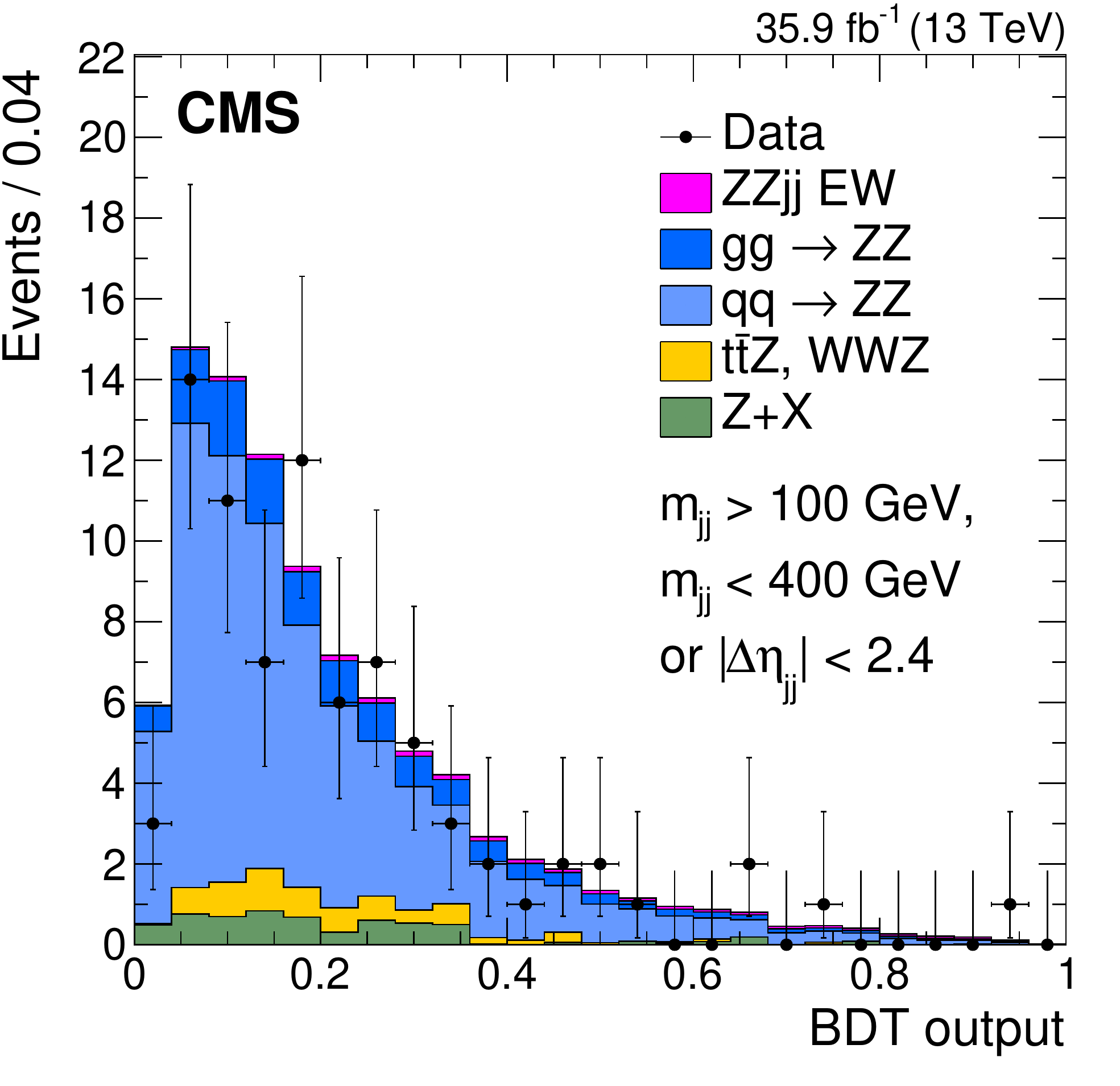}}}
\subfigure [] {\resizebox{0.4 \textwidth}{!}{\includegraphics{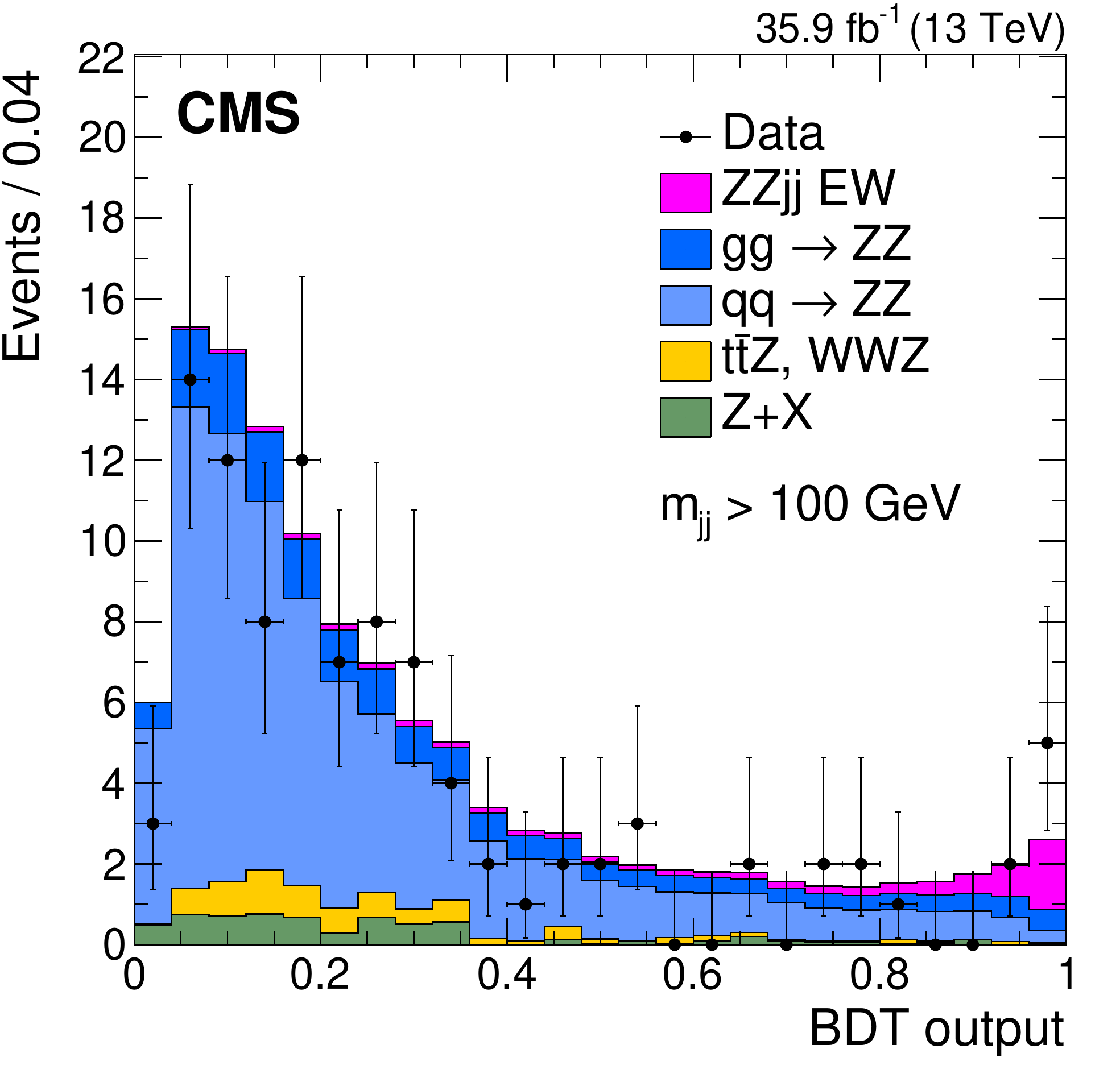}}}
\caption{Distribution of the BDT output in the control region obtained by selecting ZZjj events with $m_\mathrm{jj} < 400\GeV$ or $|\Delta\eta_\mathrm{jj}| < 2.4$ (left) and for the ZZjj selection (right)~\cite{ZZjj}. Points represent the data, filled histograms the expected signal and background contributions.
}\label{ZZjj_1}
\end{figure}

The dominant background in the measurement is the noninstrumental QCD background, which is taken from simulation. A control region enriched in this background is selected by requiring $m_\mathrm{jj} < 400\GeV$ or $|\Delta\eta_\mathrm{jj}| < 2.4$ and is used to validate the modeling provided by the simulation. Figure~\ref{ZZjj_1} (left panel) shows the good agreement between the data in the control region and the expectation for the distribution of the BDT output from the simulation.

The measurement of the signal strength $\mu$ is carried out using the BDT output distribution of all selected ZZjj events, including the QCD-enriched region. Including the later in the maximum likelihood fit permits to constrain the yield of the background to the observed data. The observed signal strength is  $\mu = 1.39^{+0.72}_{-0.57}\stat ~^{+0.46}_{-0.31}\syst = 1.39^{+0.86}_{-0.65}$, and the background-only hypothesis is excluded with a significance of $2.7$ standard deviations ($1.6$ standard deviations expected). The signal strength corresponds to a fiducial cross section of ${\sigma_{\mathrm{EW}}(\mathrm{pp}\rightarrow\mathrm{ZZjj}\rightarrow\ell\ell\ell'\ell'\mathrm{jj}) = 0.40^{+0.21}_{-0.16}\stat~^{+0.13}_{-0.09}\syst\fb}$, compatible with the SM prediction of ${0.29^{+0.02}_{-0.03}\fb}$.

The observed ZZjj events are furthermore used to constrain aQGC coupling parameters of the operators T0, T1, and T2, as well as the neutral current operators T8 and T9. The 95\% CL limits are derived from the diboson invariant mass $m_\mathrm{ZZ}$ distribution and listed in Tab.~\ref{tab:aqgc_limits}.

%Figure~\ref{ZZjj_2} shows the $m_\mathrm{ZZ}$ distribution of the data, the SM expectation, and two aQGC scenarios.
%\begin{figure}[H] \centering
%\subfigure [] {\resizebox{0.4 \textwidth}{!}{\includegraphics{aqgc_mZZ.pdf}}}
%\caption{The distribution of the diboson invariant mass in the ZZjj selection together with the SM prediction and two aQGC scenarios~\cite{ZZjj}. Points represent the data, filled histograms the expected signal and background contributions. The last bin includes all contributions with $m_\mathrm{ZZ} > 1200\GeV$.
%}\label{ZZjj_2}
%\end{figure}

\begin{table}[!h]
\vspace{0.5cm}
\begin{center}
\topcaption{Observed and expected lower and upper 95\% CL limits in units of TeV$^{-4}$.}
   \begin{tabular}{l c c c c c}
   \hline \hline
	Coupling			&  Exp. lower	& Exp. upper	& Obs.  lower	& Obs. upper	\\
\hline
        $f_{T_{0}}/\Lambda^4$       & $-0.53$	& $0.51$		& $-0.46$ 		& $0.44$  \\
        $f_{T_{1}}/\Lambda^4$       & $-0.72$	& $0.71$         	& $-0.61$		& $0.61$  \\
        $f_{T_{2}}/\Lambda^4$       & $-1.4$		& $1.4$          	& $-1.2$		& $1.2$    \\
        $f_{T_{8}}/\Lambda^4$       & $-0.99$	& $0.99$         	& $-0.84$		& $0.84$  \\
        $f_{T_{9}}/\Lambda^4$       & $-2.1$		& $2.1$          	& $-1.8$		& $1.8$    \\
  \hline
   \end{tabular}
      \label{tab:aqgc_limits}
   \end{center}
\end{table}

\section*{Conclusions}
Two recent measurements of vector boson scattering and a measurement of vector boson fusion by the CMS Collaboration have been summarized. All three measurements are based on a dataset of proton--proton collisions at $\sqrt{s}=13\TeV$ with an integrated luminosity of $35.9\fb^{-1}$. The first observation for the electroweak production of a pair of same-sign W bosons was reported with an observed (expected) significance of 5.5 (5.7) standard deviations. The first measurement of vector boson fusion of a Z boson at $\sqrt{s}=13\TeV$ was presented. A first measurement of the electroweak production of two Z bosons at the LHC was presented, reporting a signal significance of 2.7 standard deviations (1.6 standard deviations expected). The data are in general agreement with the expectations from the standard model. Limits on physics beyond the standard model were presented in the vector boson scattering analyses, providing the most stringent constraints on such scenarios to date.

%%  if necessary
%\Acknowledgements
%I am grateful to XYZ for fruitful discussions.

\end{document}